\documentclass[fleqn]{SCYE}
\usepackage{cite}
\usepackage{amsmath}
\usepackage{amssymb}

\newcommand{\aap}{    {\it Astron. Astrophys.}}
\newcommand{\aaps}{   {\it Astron. Astrophys. Suppl.}}
\newcommand{\aapr}{   {\it Astron. Astrophys. Rev.}}

\newcommand{\apj}{    {\it Astrophys. J.}}
\newcommand{\apjl}{   {\it Astrophys. J. Lett.}}
\newcommand{\araa}{   {\it Ann. Rev. Astron. and Astrophys}}
\newcommand{\apjs}{   {\it Astrophys. J. Suppl. Ser.}}

\newcommand{\grl}{    {\it Geophys. Res. Lett.}}

\newcommand{\jgr}{    {\it J. Geophys. Res.}}
\newcommand{\mnras}{  {\it Mon. Not. Roy. Astron. Soc.}}

\newcommand{\solphys}{{\it Solar Phys.}}

\newcommand{\ssr}{    {\it Space Sci. Rev.}}

\newcommand{\zap}{    {\it Z. Astrophys.}}
\newcommand{\planss}{    {\it Planetary and Space Science}}
\newcommand{\cjss}{    {\it Chin. J. Space Sci.}}

\begin{document}

\ArticleType{Review}
\Year{2020}
\Month{May}
\Vol{***}
\No{***}
\DOI{***}
\BeginPage{***} 
\EndPage{***}
\ReceiveDate{May, 2020}
\AcceptDate{June, 2020}

\title{Characteristics and applications of interplanetary coronal mass ejection composition}
{Characteristics and applications of interplanetary coronal mass ejection composition}

\author[1]{SONG Hongqiang}{hqsong@sdu.edu.cn}%
\author[2]{YAO Shuo}{}

\AuthorMark{Song H Q, Yao S}

\AuthorCitation{Song H Q, Yao S}

\address[1]{Shandong Provincial Key Laboratory of Optical Astronomy and Solar-Terrestrial Environment,\\ and Institute of Space Sciences, Shandong University, Weihai, Shandong {\rm 264209}, China}
\address[2]{School of Geophysics and Information Technology, China University of Geosciences (Beijing), Beijing {\rm 100083}, China}


\abstract{In situ measurements of interplanetary coronal mass ejection (ICME) composition, including elemental abundances and charge states of heavy ions, open a new avenue to study coronal mass ejections (CMEs) besides remote-sensing observations. The ratios between different elemental abundances can diagnose the plasma origin of CMEs (e.g., from the corona or chromosphere/photosphere) due to the first ionization potential (FIP) effect, which means elements with different FIP get fractionated between the photosphere and corona. The ratios between different charge states of a specific element can provide the electron temperature of CMEs in the corona due to the freeze-in effect, which can be used to investigate their eruption process. In this review, we first give an overview of the ICME composition and then demonstrate their applications in investigating some important subjects related to CMEs, such as the origin of filament plasma and the eruption process of magnetic flux ropes. Finally, we point out several important questions that should be addressed further for better utilizing the ICME composition to study CMEs.}

\keywords{coronal mass ejection, interplanetary coronal mass ejection, elemental abundance, ionic charge state, magnetic flux rope, magnetic cloud, filament, flare}

\maketitle



\begin{multicols}{2}

\section{Introduction}\label{section1}
Interplanetary coronal mass ejections (ICMEs) refer to the counterpart of coronal mass ejections (CMEs) in the interplanetary space, which are an energetic explosive phenomenon occurred in the solar atmosphere \cite{chen2011CME,webb2012CME,cheng2017,guo2017}. When interacting with the Earth's atmosphere, ICMEs can induce strong geomagnetic activity \cite{gosling1991gm,xumengjiao2019} and seriously influence our high-technology activities through damaging satellites, overloading power grids, and disrupting GPS navigation systems \cite{cannon2013,riley2018}. Therefore, it is of great significance to investigate CMEs/ICMEs for both astrophysics and space weather.

Theoretical studies suggest that CMEs take place due to the eruption of magnetic flux rope (MFR, a coherently \Authorfootnote helical magnetic structure with its field lines winding  around one central axis more than one turn), which can be formed prior to \cite{lin2000,patsourakos2013,wangwensi2017} or during solar eruptions \cite{mikic1994,song2014a,ouyang2015} through magnetic reconnection occurred in the corona. So far, none of physical mechanisms can produce a CME without involving the MFR. In the case that the MFR has existed in the corona before eruption, theoretical studies propose an alternative mechanism to answer where and how the MFR is built up, which suggests that the MFR is formed in the convection zone and can emerge into the corona by buoyancy \cite{fan2001,magara2004,leake2013}. This mechanism is supported by some observations \cite{okamoto2008}. However, simulations found that only the upper part of the MFR can emerge into the corona \cite{manchester2004} and the reconnection is necessary to transfer some emerged magnetic fluxes into a new MFR structure in the corona \cite{leake2014}.

Observational studies demonstrate that CMEs often result from the eruption of filaments \cite{webb1987}, coronal cavities \cite{gibson2006}, sigmoids \cite{titov1999}, or hot channels \cite{zhangjie2012nc}. Filaments are cold and dense plasmas usually located at the MFR dips with the magnetic field lines being horizontal locally and curved upward \cite{kippenhahn1957}, and they will be termed as prominences when appeared around the solar limb. Alternatively, filaments can also be supported by the sheard magnetic arcades \cite{chen2014}. Coronal cavities refer to the elliptical regions with lower density and usually locate above and around the prominences, which can be observed in a variety of wavelength, such as white light, radio, soft X-ray, as well as extreme ultraviolet (EUV). Sigmoids appear as the forward or inverse S-shaped structure usually in active regions. Hot channels (hot blobs when observing along the axis due to the projection effect) are one high-temperature coronal structure observed in the 131 \AA\ or 94 \AA\ passbands. The erupted sigmoid can appear as a hot channel \cite{liurui2010}. Many studies support that the coronal cavities \cite{wang2008}, sigmoids \cite{cheng2015}, and hot channels \cite{song2015evidence} are indicative of MFRs. It is natural that the MFR eruptions in theory manifest as the ejections of filaments/prominences, coronal cavities, sigmoids, or hot channels in observations. See a recent review \cite{cheng2017} for more details. Therefore, both the origin of filament plasma \cite{song2017origin} and the formation of MFRs \cite{song2016} are important issues for CMEs, which will be discussed more in the following sections.

Since the first observation of CMEs in the early 1970s, many observational studies about CMEs near the Sun have been conducted through remote-sensing data at various passbands, including the X-ray, EUV, white light, as well as radio \cite{chen2011CME}. The white-light coronagraphs, e.g., the Large Angle Spectrometric Coronagraph (LASCO) on board the Solar and Heliospheric Observatory (SOHO) and CORs onboard the Solar TErrestrial RElations Observatory (STEREO), provide many properties of CMEs near the Sun, including the structure, volume, mass, velocity, acceleration, as well as occurrence rate \cite{webb2012CME}. The solar spectroscopy can diagnose the temperature, density, line-of-sight velocity, and composition related to CMEs \cite{ciaravella2002,tian2011,tian2012}. Also, the temperature and density of CMEs can be acquired through differential emission measure (DEM) \cite{cheng2012}. All of the knowledge strongly improves our understanding on the initiation and acceleration mechanisms of CMEs.

However, the density inversion based on white-light brightness critically depends on the 3D structure of CMEs and the removal of the F corona \cite{hayes2001}. Spectroscopic observations and DEM require sophisticated models to get the physical parameters, and the models rely on the atomic and molecular data measured in laboratory or calculated from atomic physics. Thus the inferred parameters could vary with time. For instance, the elemental abundances derived from spectroscopic data have changed substantially over time as the development of inversion techniques \cite{zurbuchen2016}. In addition, the remote-sensing observations are usually influenced by the projection effect, and they can not provide the reliable magnetic field measurements of CMEs either. Partial disadvantages of remote-sensing observations can be overcome by the in situ measurements of ICMEs, especially for the composition of CMEs. In the meantime, the in situ measurements of magnetic field within ICMEs directly prove that CMEs can contain the MFR structure \cite{burlaga1981,burlaga1988,liuying2008}.

So far, the most complete composition data of interplanetary plasma is provided by the solar wind ion composition spectrometer (SWICS) on board both the Ulysses \cite{gloeckler1992} and the Advanced Composition Explorer (ACE) \cite{gloeckler1998}. The solar wind particles collected by SWICS are analyzed on board and the data are telemetered to the Earth. The SWICS is optimized to acquire the low-noise measurements of elemental and charge-state composition in the interplanetary space. The particles detected by SWICS cover an energy-per-charge (E/Q) range of 0.49--100.0 keV e$^{-1}$ and all of the masses from H to Fe \cite{gilbert2012}. The instrument can measure the speed, mass, and charge of incident particles through three techniques, i.e., the electrostatic analysis, the time-of-flight analysis, as well as the total energy measurement \cite{gilbert2014,zurbuchen2016}. SWICS can measure the charge-state distributions and abundances of $\sim$10 elements \cite{lepri2001}. The newly released SWICS/ACE data are calculated through a completed statistically accurate inversion model \cite{shearer2014}, which takes into account both systematic and statistical uncertainties and results in an increasing identification of rare ions such as Fe$^{6+}$ and Fe$^{7+}$. This is very crucial for identifying the cold filament plasma in the interplanetary space \cite{song2017origin,wangjiemin2018,fengxuedong2018,lidongni2020}.

In addition to ICMEs, the non-transient solar wind in situ composition also provides us with a powerful tool to investigate their mechanisms of origin, heating and acceleration \cite{geiss1995,chenyao2004a,chenyao2004b,lepri2013,zhaoliang2014,zhaoliang2017,Fuhui2017,Fuhui2018}. In the current review, we focus on the ICME composition observations and their applications on CME studies. This paper is organized as follows: in Section 2, we summarize the criteria to identify ICMEs from the background solar wind. Section 3 introduces the statistical results of ICME elemental abundances and the application to identify the origin of filament plasma through abundances. Section 4 presents the observational results of ICME ionic charge states and the application to analyze the MFR formation process through charge states. Section 5 is our summary and discussion.

\section{Identification of ICMEs}\label{section1}
Many studies have been conducted on ICMEs \cite{cane2003,gopalswamy2006,lepping2015a,wucc2015,chiyutian2016,jianlan2018} and their propagation from the Sun to the Earth \cite{cargill2004,howard2006,shenfang2011,liuying2014,hess2015,chenchong2019} in recent decades. The twin STEREO spacecraft enables us to track ICMEs continuously in the heliosphere through a geometric triangulation \cite{liuying2010}. Based on their magnetic field and plasma measurements, ICMEs are divided into two categories: magnetic clouds (MCs, \cite{burlaga1981}) that contain the regular MFR structure, and non-cloud ICMEs that have irregular magnetic field features. Gosling \cite{gosling1990} reported that $\sim$30\% of ICMEs in 1978-1982 were MCs, while Marubashi \cite{marubashi2000} claimed that up to $\sim$80\% of ICMEs were MFR encounters. There also exists a solar cycle effect for MC fraction in ICMEs, ranging from $\sim$15\% near solar maximum to $>$60\% near solar minimum \cite{cane2003}. It is unclear whether these two categories result from the different CME initial mechanisms in the corona, the complex propagation process in the interplanetary space, or just different spatial sampling of ICMEs \cite{riley2006}.

In the previous and current researches, many characteristics have been invoked to identify ICMEs in the background solar wind \cite{zurbuchen2006,wucc2011,richardson1995,liuying2005}, which can be separated into magnetic field, plasma dynamics, suprathermal particle, plasma composition, and plasma wave signatures \cite{zurbuchen2006}. The potential signatures of ICMEs include 1) enhanced magnetic field strength, 2) field direction changing smoothly, 3) low proton temperature, 4) low plasma beta, 5) enhanced helium abundance, 6) bidirectional streaming of super-thermal electrons, 7) bidirectional streaming of low energy protons, 8) high charge states related to magnetic reconnection, 9) low charge states related to filaments, 10) singly charged helium (He$^{+}$), 11) bidirectional particle flows at cosmic ray energies (1 MeV), 12) bidirectional solar wind electron heat flux events, 13) the occurrence of Forbush decrease, and so on \cite{wucc2011,liuying2005}. Usually the first four signatures are used to identify MCs. The elemental abundances and charge states of heavy ions can vary significantly case by case, thus usually they are not taken as the main criteria to identify ICMEs. Especially, they can not be used to determine the boundaries of ICMEs due to their diversified profiles within ICMEs as presented in Section 4.

Generally, it is rare for a single ICME event to possess all of the above 13 characteristics, and various signatures might not occur simultaneously. This is not surprising since they arise from different physical circumstances. For example, the appearance of low charge states may depend on whether the filament plasma is detected, and the bidirectional streaming of electrons relies on the magnetic field line connectivity to the Sun. In addition, no magnetic field rotation signature can be detected if the observing spacecraft makes a glancing encounter with the MC. The practical identification of ICMEs is to examine as more criteria as possible and reach a consensus based on several criteria during a potential interval of the solar wind \cite{zurbuchen2006}.

Most MC parameters, such as the temperature, density, velocity, morphology, volume, as well as magnetic field strength and structure, experience large variation during their transit from the Sun to 1 AU due to the expansion, acceleration/deceleration, and interaction with background solar wind or other large scale interplanetary structures. However, the elemental abundances and ionic charge states keep unchanged during the transit, which provide us an excellent opportunity to investigate some important issues related to CMEs.

\section{Elemental abundances of ICMEs}\label{section1}
\subsection{The FIP effect }\label{section1}
The elemental abundances of the solar corona and the photosphere are different because elements with different first ionization potential (FIP) get fractionated between the photosphere and corona (FIP effect \cite{laming2009}). The FIP fractionation should locate in the chromosphere \cite{geiss1982}, in which low-FIP ($<$ 10 eV) elements are generally ionized and high-FIP ($>$ 10 eV) elements are partially neutral at least. Researchers have proposed several models to explain the FIP effect \cite{laming2004,laming2009}. The early theoretical models are correlated with the diffusion, thermoelectric driving, chromospheric reconnection, or ion cyclotron wave heating \cite{laming2015}. Laming proposed a model correlated with ponderomotive forces, which are the time-averaged nonlinear forces acting on a media in the presence of oscillating electromagnetic fields \cite{lundin2006}. The model demonstrated that the elemental fractionation could be generated by the ponderomotive force in the chromosphere from Alfv\'{e}n waves, which is directed upward usually, and acts only on chromospheric ions, not neutrals \cite{laming2004}. Therefore, the low-FIP elements, which are predominantly ionized in the chromosphere, are enhanced in abundance when they flow into the corona, and the high-FIP elements, which are largely neutral, appear essentially unaffected \cite{laming2009}.

The degree of chemical separation varies obviously in different regions. In the coronal quiet region and slow solar wind, the elements with low-FIP are enhanced in abundance by a factor of $\sim$3, typically ranging from 2 to 5, which is established by both in situ and remote-sensing measurements \cite{laming2015}. In the coronal holes and fast solar wind, the degree of FIP fractionation is significantly smaller than that in the quiet region and slow wind \cite{bochsler2007,feldman1998,laming2015}. A recent study confirmed that the FIP bias factors of those low-FIP elements are enhanced more in winds originated from the hot coronal regions than in winds from the cold coronal regions, and found that the FIP bias factors can also be affected by the solar cycle \cite{zhaoliang2017apj}. Active regions and flares also have different elemental fractionation, usually reduced from that measured in the quiet region \cite{laming2015}. Note that the abundances of high mass-to-charge ions in solar energetic particles are greatly enhanced compared with coronal values \cite{reames1994}, which may result from the mechanisms of ion acceleration during flare magnetic reconnection \cite{drake2009}.

In addition, Widing and Feldman \cite{widing2001} found that the new coronal loops emerged with photospheric abundances, while gradually changed to coronal abundances over the course of a few days, which means the elemental abundances could be different between nascent and mature active regions. Consequently, CMEs originated from different coronal regions (quiet region, nascent or mature active region) can have different elemental abundances. Besides, the elemental abundance should not be uniform everywhere within an ICME (e.g., containing the filament or not), and the measured result also depends on where the spacecraft passes through a particular ICME. Therefore, the observed elemental abundances of ICMEs are highly variable \cite{reisenfeld2007} from event to event, and we mainly introduce the statistical result on the ICME elemental abundances in Section 3.2.

\subsection{The statistical result}\label{section1}
The elemental abundances of ICMEs are of interest due to their diagnostic values for analyzing the origin of ICME-related plasmas that are released from the magnetically closed region. Recently, Zurbuchen et al. \cite{zurbuchen2016} conducted a comprehensive statistics on the abundances of 310 ICMEs in Richardson \& Cane's list from 1998 March to 2011 August using the newly released data that include the inversion of low-abundance elements such as Ne, S, and Si, and compared them to both slow and fast solar wind. They first identify the ``compositionally hot" ICMEs using the methodology proposed by Lepri et al. \cite{lepri2001}, which regarded an ICME as ``compositionally hot" when it contained at least 6 hr of plasma with average Fe charge state $<$Q$_{Fe}$$>$ beyond 12+. With this methodology, 47.7\% of the ICMEs are identified as ``compositionally hot", which are dubbed as high $<$Q$_{Fe}$$>$ ICMEs.

To present the general behavior of the average elemental abundances of ICMEs, they analyzed the Mg/O, Si/O, and Ne/O as functions of Fe/O for two classes of ICMEs as shown in Figure 1, where the high $<$Q$_{Fe}$$>$ ICMEs are plotted in orange, and the other ICMEs in black \cite{zurbuchen2016}. The fitted line in this figure are linear in log-space, see Table 1 in reference \cite{zurbuchen2016} for the fitting parameters. Figure 1 shows that the range of compositional distribution for each compositional quantity extends over one order of magnitude. The positive correlation of Mg/O and Si/O with respect to Fe/O demonstrate that the FIP effect affects the metal-like ions in a similar fashion. The strongest correlations exist between Si/O and Fe/O, which could be understood as the FIPs of Mg, Si, and Fe are 7.65, 8.2, and 7.9 eV, respectively. However, it is surprising there exists a strong positive correlation of Ne/O for high $<$Q$_{Fe}$$>$ ICMEs as shown in Figure 1(c), as the FIP of Ne is high (21.6 eV) \cite{zurbuchen2016}. Their results clearly show that the elemental correlation is stronger in each case for high $<$Q$_{Fe}$$>$ ICMEs and weak for the other ICMEs, especially for Ne. They concluded that high $<$Q$_{Fe}$$>$ ICMEs have an obviously different elemental composition compared to the other ICMEs \cite{zurbuchen2016}.

\begin{figure}[H]
\centering
\includegraphics[width=0.48\textwidth]{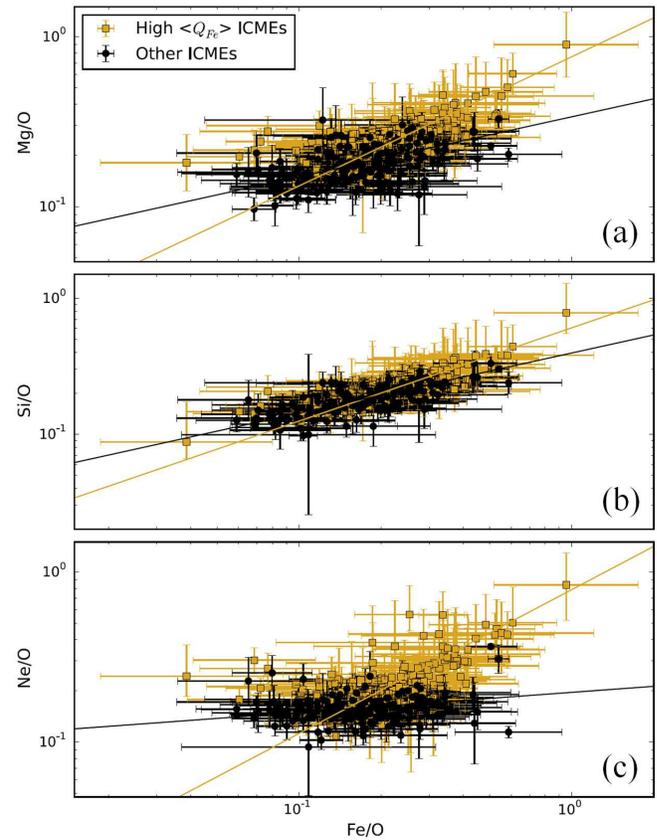}
\caption{Elemental composition of both high $<$Q$_{Fe}$$>$ ICMEs and the other ICMEs. The error bar indicates the standard deviation of a compositional quantity within the ICME interval identified by Richardson and Cane. For each element, the slope of high $<$Q$_{Fe}$$>$ ICMEs is steeper compared to that of the other ICMEs \cite{zurbuchen2016}.}
\label{Fig1}
\end{figure}

To investigate the difference between the two ICME populations quantitatively, Zurbuchen et al. \cite{zurbuchen2016} computed the average composition for each ICME population and compared the values to those of both slow and fast solar wind. The relative abundances (relative to O) are shown in Figure 2(a). Clearly, there exist significant increasing factors of FIP enhancement. For low-FIP elements such as Mg, Fe, and Si, their largest FIP enhancements exist in the high $<$Q$_{Fe}$$>$ ICMEs, and their smallest FIP factors are seen for fast wind. The other ICMEs and slow wind are intermediate between the two extremes \cite{zurbuchen2016}.

\begin{figure}[H]
\centering
\includegraphics[width=0.48\textwidth]{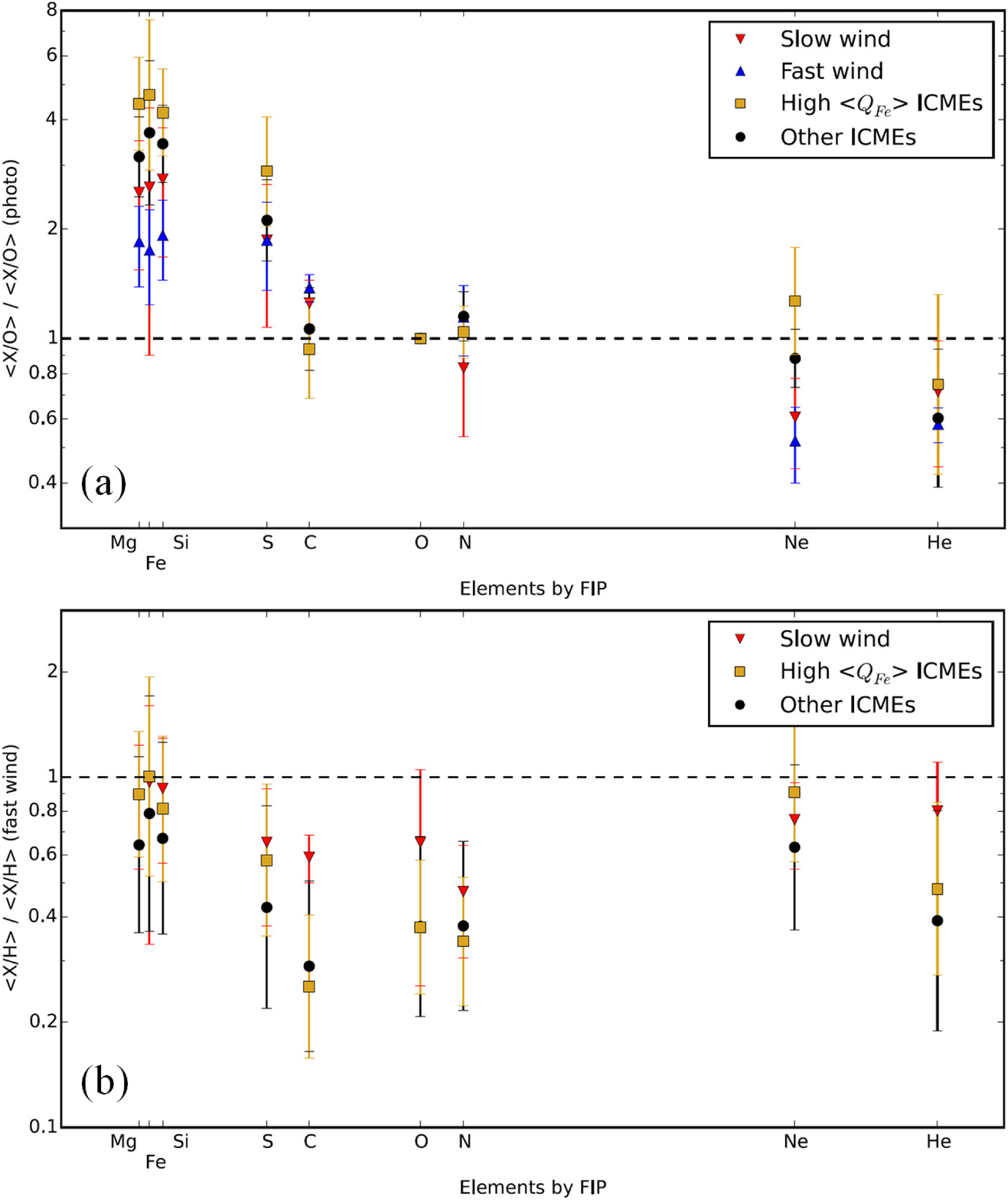}
\caption{(a) The relative abundances of high $<$Q$_{Fe}$$>$ ICMEs and the other ICMEs compared to those of fast and slow solar wind. The high $<$Q$_{Fe}$$>$ ICMEs exhibit a higher FIP than both the other ICMEs and slow wind. The Ne/O composition of high $<$Q$_{Fe}$$>$ ICMEs exceeds that of solar wind. (b) The absolute abundances of slow solar wind and two types of ICMEs compared to those of fast wind, which is the most photosphere-like sample available to us in the interplanetary space. High $<$Q$_{Fe}$$>$ ICMEs are depleted for S, C, O, and He, but are identical to those of polar coronal holes for Mg, Fe, Si, and Ne \cite{zurbuchen2016}.}
\label{Fig2}
\end{figure}

The relative abundances of mid-FIP elements (S, C, and N) exhibit different behaviors \cite{zurbuchen2016}. For example, S/O presents enhancements with the ordering similar to the metals, while its enhancements are smaller. The C/O abundances of both type ICMEs show a depletion compared to average abundances of both types of solar wind. The depletions are relatively small, and there is no significant difference between high $<$Q$_{Fe}$$>$ ICMEs and the other ICMEs for C/O. The N/O abundance of ICMEs exhibit an insignificant enhancement over the solar wind values \cite{zurbuchen2016}.

The relative abundances of high-FIP elements, such as the noble elements Ne and He, possess the rather peculiar behaviors \cite{zurbuchen2016}. The high $<$Q$_{Fe}$$>$ ICMEs present an enhancement in Ne/O, exceeding the solar wind by a factor $\sim$2. The other ICMEs have a small elevated abundance compared to slow wind, while it is obviously lower than those of high $<$Q$_{Fe}$$>$ ICMEs. The He/O abundances of both ICMEs and solar wind show the depletion as compared to the photospheric value \cite{zurbuchen2016}.

Zurbuchen et al. \cite{zurbuchen2016} summarized that the composition of the other ICMEs is intermediate between those of the high $<$Q$_{Fe}$$>$ ICMEs and slow wind, which can indicate their actual intermediate states, and might also indicate the cross-contamination of both compositions. To determine the physical interpretation for the results, they further analyzed the absolute abundances of ICMEs and solar wind as shown in Figure 2(b) by using H/O values from \cite{steiger2010}. The abundances in Figure 2(b) are normalized by the abundance of fast solar wind, which exhibits only a small FIP effect \cite{zurbuchen2016}.

The abundances of low-FIP elements (Mg/H, Fe/H, and Si/H) of both ICMEs and slow wind are identical to that of fast wind within the error bars. Similarly, Ne/H for high $<$Q$_{Fe}$$>$ ICMEs is consistent with that of fast wind and somewhat reduced in both the other ICMEs and the slow wind. Middle-FIP elements (S/H, C/H, N/H, and O/H) are reduced in all samples, and all their values are depleted more within ICME samples compared to the slow wind. The He/H abundance presents an overall depletion compared to the solar wind though there exist cases of enhanced He/H in some ICMEs \cite{zurbuchen2006}. The absolute abundance of low-FIP elements keeps approximately constant for all of the samples, and the differences between the low- and high-FIP elements are obvious, which support that the anomalies of coronal elemental abundances result from the depletion of high-FIP and mid-FIP elements, instead of the enrichment of elements with low-FIPs \cite{bochsler2000}.

Except comparing the elemental abundance of the high $<$Q$_{Fe}$$>$ ICMEs with that of the other ICMEs, researchers also made a comparison between the MCs and non-cloud ICMEs \cite{owens2018}, which showed that the Fe/O is obviously elevated in MCs (especially the fast events with velocity beyond 450 km s$^{-1}$) compared to the non-cloud ICMEs, see the Figures 5 and 8 in \cite{owens2018}.

We mainly described the overall situation about abundances of heavy elements that are rich relatively in the solar atmosphere. The studies on the other elements with much less abundances as compared to O, C, Fe and so on are relatively rare, as it is difficult to measure the elements with low abundances. Giammanco et al. \cite{giammanco2008} analyzed 4 years of CELIAS (Charge, Element, and Isotope Analysis System) data \cite{hovestadt1995} and reported the measurements of the absolute abundances of Al, Na, P, and K in the solar wind for the first time, as well as their elemental ratios relative to Ca and Mg. They gave the elemental enrichment as a function of the FIP and the first ionization time based on their abundance measurements.

\subsection{The origin of filament plasma}\label{section1}
 \begin{figure}[H]
 \centering
 \includegraphics[width=0.48\textwidth]{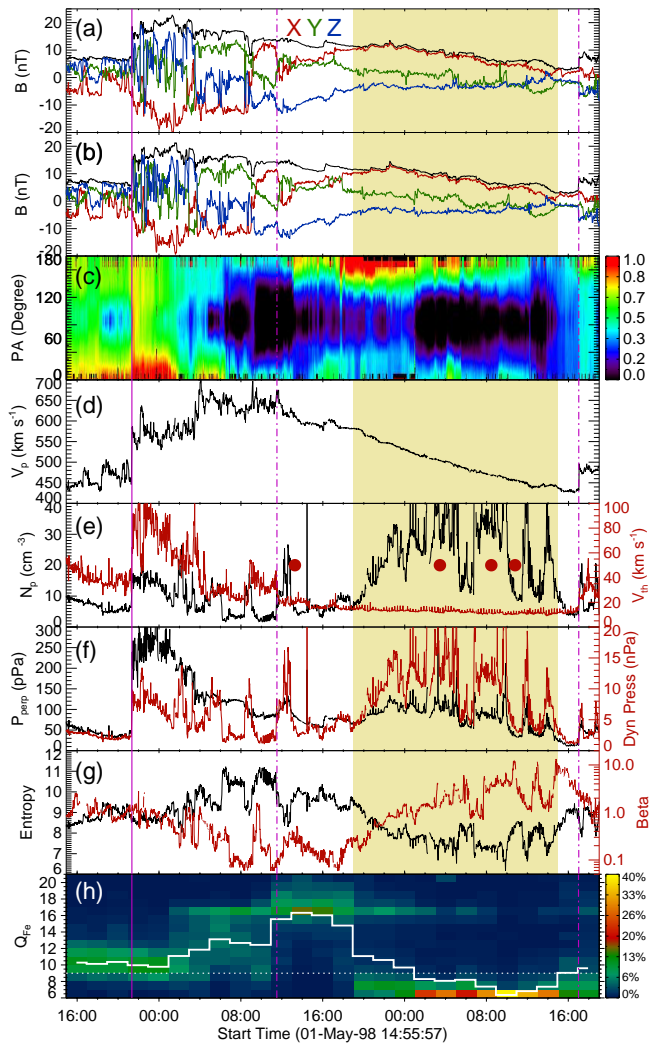}
 \caption{Magnetic field and plasma parameters measured by WIND and ACE. The panels from top to bottom display the total magnetic field strength (black) and its three components provided by ACE (a) and WIND (b), PAD of super-thermal electrons (c), bulk speed (d), density (black) and thermal velocity (red) (e), the total pressure along the perpendicular direction (black) and the dynamic pressure (red) (f), entropy (black) and plasma $\beta$ (red) (g), charge-state-distribution map of Fe ions and the $<$Q$_{Fe}$$>$ (white) (h). The vertical solid line and the two vertical dotted-dashed lines denote the ICME shock and two boundaries of the MC, respectively. The shaded region corresponds to the filament interval \cite{song2017origin}.}
 \label{Fig3}
 \end{figure}
In this subsection, we demonstrate one application of in situ elemental abundance on the CME-related study. It has been reported that more than 70\% of CMEs are correlated with the filament eruptions \cite{gopalswamy2003}, thus it is important to address where the filament plasma comes from. Two types of models have been proposed: one claims that the filament plasma is from solar photosphere and/or chromosphere through siphon effect and/or evaporation as well as injection process \cite{spicer1998,mackay2010}; the other one proposes that the coronal plasma can condense and form the filament directly due to the thermal instability \cite{sakai1987,demoulin1993}. The siphon and evaporation/injection models predict that the filament abundances are more close to the photospheric ones, while the condensation models would have the coronal abundances. Therefore, we can constrain the models through the elemental abundance analysis.

Some abundance analyses through spectroscopy demonstrated that the Mg/Ne ratios of prominences have values intermediate between the corona and photosphere but none are as high as those in the corona \cite{spicer1998}, which does not support the condensation models. Recently, Parenti et al. \cite{parenti2019} used the DEM technique to investigate the elemental abundances of two prominences, and also showed that they had the photospheric abundances, consistent with the spectroscopy results. However, the results from spectroscopy and DEM depend on the physical models as mentioned. Besides, there is a particular problem for the prominence analysis as it is optically thick. The radiation transfer process hinders researchers to derive the precise abundances within the prominence interior. To further address the origin of filament plasma, one method is to measure the elemental abundances of filaments in the interplanetary space. This requires to identify the the filament within ICMEs clearly first. The filament is cooler and denser as compared to the corona, thus the most profound signature of filament plasma is the appearance of a significant fraction of low charge ions, e.g., Fe$^{6+}$ and even more lower \cite{lepri2010,gruesbeck2012}, instead of Fe$^{10+}$ in the solar wind or Fe$^{16+}$ within the hot intervals of ICMEs \cite{song2016}.

Song et al. \cite{song2017origin} reported such a good in situ detection of filament. A halo CME resulted from a filament eruption took place on 1998 April 29, and the corresponding ICME was detected by both WIND and ACE at L1 point. Skoug et al. observed an enhancement of He$^+$ during the latter half of this ICME \cite{skoug1999}, and Gloeckler et al. found a wide variety of both unusually high and unusually low ionic charge states in this event \cite{gloeckler1999}. This event provides us a precious opportunity to study the origin of filament plasma as it contains long ($\sim$20 hr) and obvious signs of filament as shown in Figure 3 \cite{song2017origin}. Usually the detected filament durations within ICMEs last only one to several hours \cite{yaoshuo2010,fengxuedong2018}, which is unsuitable for the analysis of elemental abundances as the typical temporal resolution of SWICS is 2 hr.

In Figure 3, the ICME shock is depicted with the purple vertical solid line, and the MC boundaries two purple vertical dotted-dashed lines. Figures 3(a) and 3(b) show the magnetic field measured by the ACE and WIND, respectively, including the total field strength (black) and the three components (X, Y, and Z in red, green, and blue, respectively) in the Geocentric Solar Ecliptic coordinate. The magnetic field profiles are almost identical in the two panels, which proves that both ACE and WIND measured the same part of the ICME. Figure 3(c) presents the normalized pitch angle distribution (PAD) of super-thermal electrons (272 eV), exhibiting the bidirectional electrons during the MC passages. Figure 3(d) shows that the velocity within the MC decreases gradually with time, agreeing with the expectation of the MFR expansion. Figure 3(e) presents the number density (black) and the thermal velocity (red) of protons. The total perpendicular (black) and the dynamical (red) pressures are presented in Figure 3(f). Figure 3(g) exhibits the Entropy (black) and the plasma $\beta$ (red). The last panel is the charge-state-distribution map of Fe ions and the profile of $<$Q$_{Fe}$$>$ (white) with a temporal resolution being 2 hr \cite{song2017origin}. The MC interval contains an obvious low Fe charge states and high proton density as indicated with the yellow shade. This agrees with the filament features, strongly proving that the yellow region corresponds to the the erupted filament. Song et al. \cite{song2017origin} fitted this MC through a velocity-modified Gold and Hoyle model \cite{wangyuming2016a} and found the spacecraft trajectory was close to the MC center. This might be one reason why the filament interval was so long.

\begin{figure}[H]
 \centering
 \includegraphics[width=0.48\textwidth]{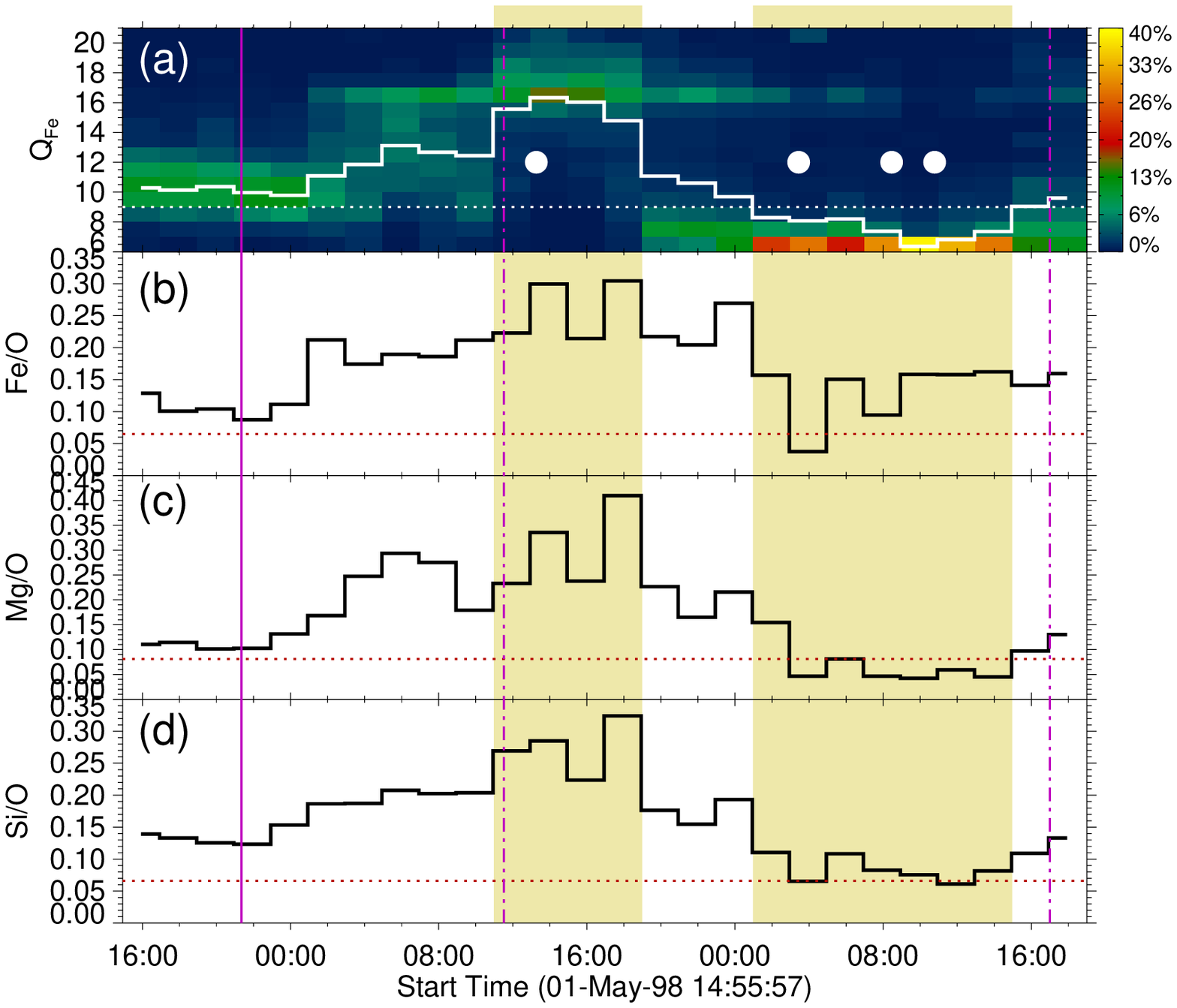}
 \caption{Charge states and relative abundances measured by ACE. The panels sequentially present the charge-state-distribution-map of Fe ions and the profile of $<$Q$_{Fe}$$>$ (a), the abundances of Fe/O, Mg/O, and Si/O (b--d). The corresponding abundance ratios in the photosphere are denoted with the red horizontal dotted lines. The vertical solid line and the two vertical dotted-dashed lines denote the ICME shock and two boundaries of the MC, respectively. The left (right) shaded region corresponds to the non-filament (filament) interval} \cite{song2017origin}.
 \label{Fig4}
 \end{figure}

To infer the filament plasma origin, Song et al. \cite{song2017origin} analyzed the elemental abundances as shown in Figure 4, where the charge-state-distribution map of Fe ions are re-plotted in Figure 4(a) to show the filament interval clearly. Two regions with higher and lower Fe charge states are emphasized with yellow shades, corresponding to the non-filament (left) and filament (right) intervals within the ICME, respectively. Figures 4(b)-(d) present the Fe/O, Mg/O, and Si/O ratios sequentially, with the horizontal dotted lines in each panel depicting the corresponding abundances in the photosphere \cite{asplund2009}. It is obvious that the two shaded regions have different ratios. The Fe/O (Mg/O and Si/O) abundances in the non-filament region are $\sim$0.25 (0.30 and 0.30), higher than the corresponding value 0.15 (0.05 and 0.07) in the filament region, which are closer to the photospheric abundances of 0.065 (0.081 and 0.066). Thus the in situ observations do not support the condensation models either. The remote-sensing imaging observations further demonstrate that the jets occurring in the vicinity of the filament footpoint can eject plasma to supply material for the filament \cite{wangjincheng2018}. Conversely, the non-filament region exhibits obvious FIP effect, which indicates that the MFR is mainly built up in the corona, instead of emerging into the corona from the convection zone.

\section{Ionic charge states of ICMEs}\label{section1}
\subsection{Freeze-in of ionic charge states}\label{section1}
The ionic charge states of ICMEs provide us another clue to study CME-related issues, e.g., the MFR formation and eruption mechanisms, due to the so-called freeze-in effect. When the MFR erupts and propagates outward, its electron density decreases rapidly with solar distance. As the ionization and recombination rates are proportional to the electron density, they also decrease with distance and shut down when the electron density is low enough. This cause the ionic charge states of plasma to freeze-in \cite{owocki1983,zhaoliang2017}. Charge states of carbon and oxygen in the solar wind usually freeze-in around 1--1.5 solar radii \cite{chenyao2003,landi2012}, and iron below 3--4 solar radii \cite{buergi1986,boe2018}. The ionic charge states measured in ICMEs beyond the freeze-in height maintain the thermal properties of the CME plasma below their freeze-in height, and they can be used to infer the eruption process of MFRs.

Generally, the high ionic charge states imply high freeze-in temperatures \cite{owens2018} though both the electron density and plasma velocity also play a role in the freeze-in process \cite{landi2012,zhaoliang2014}. The freeze-in process in CMEs can occur at different freeze-in temperatures from event to event, mainly depending on the temperature of current sheet beneath CMEs \cite{lin2000}, which means the charge states of ICMEs could be higher or normal as compared to the solar wind. Meanwhile, the ICMEs can exhibit charge states lower than that of solar wind when the spacecraft detects filament plasma \cite{wangjiemin2018,fengxuedong2018,lidongni2020}. In addition, the measured results of ICME charge states also depend on where the spacecraft crosses a particular ICME. Therefore, the observed charge states of ICMEs are also highly variable from case to case. We will describe a typical case first in Section 4.2 and then introduce a statistical result about the C, O, as well as Fe charge states within ICMEs in Section 4.3.

\subsection{The typical ionic charge states of both cold and hot ICME components}\label{section1}
Most heavy elements of ICMEs and solar wind are presented in a number of different charge states. The ratio of any two charge states of an element can be converted to a freeze-in temperature. Therefore, the charge state ratios of heavy ions act as the coronal thermometer. The dominant solar wind ionic charge states are C$^{5+}$, O$^{6+}$, and Fe$^{10+}$, which freeze-in near coronal electron temperature of 1 MK. More than 80\% of ICMEs possess the elevated charge states, e.g., O$^{8+}$, O$^{7+}$, C$^{6+}$, and Fe$^{16+}$ \cite{lepri2001,jianlan2018}, and Lepri et al. demonstrated that the high Fe charge states ($\geqslant $ 16+) can act as a good identifier of ICMEs \cite{lepri2001}. Besides, high oxygen charge states (O$^{7+}$/O$^{6+}$ $>$ 1) are also associated with many ICMEs \cite{henke1998}. However, some ICMEs can also exhibit cold component with lower charge states as mentioned \cite{wangjiemin2018,fengxuedong2018,lidongni2020}.

Lepri et al. \cite{lepri2010} presented a systematic search for cold material in ICMEs. They used a novel technique that takes advantage of the full power of SWICS on board the ACE, allowing them to analyze low-charge-state ions such as C$^{2+}$, O$^{2+}$, and Fe$^{4+}$. Three examples of charge state distributions for C, O, and Fe are presented in Figure 5. The hot and cold ICME components are shown as red and blue bars in the three panels, and the normal solar wind is shown as green bars. The charge state distributions of normal solar wind peak at C$^{5+}$, O$^{6+}$, and Fe$^{10+}$, implying the freezing temperatures around 1 MK. The hot ICME component contains significant contributions of C$^{6+}$, O$^{7+}$, Fe$^{16+}$ and beyond. The cold ICME component exhibits strong peaks at the low charge states such as C$^{2+}$, O$^{2+}$, and Fe$^{6+}$. The cold ICME component might extend to lower charge states that are not covered by their analysis method \cite{lepri2010}, which is confirmed by Gilbert et al. \cite{gilbert2012}.

In addition, Figure 5 shows that there exist contemporaneous measurements of high charge stages in this cold ICME component such as C$^{4-6+}$, O$^{6-7+}$, and Fe$^{12-17+}$, which might be caused by the mixing of plasma with different temperatures \cite{gruesbeck2012}. Therefore, we need caution to analyze the elemental abundances of filament plasma through the in situ measurements. It should be conducted when the high charge charge stages are rare, i.e., the hot component is negligible like the case shown in Figure 4.

\begin{figure*}[t]
\centering
\includegraphics[width=0.95\textwidth]{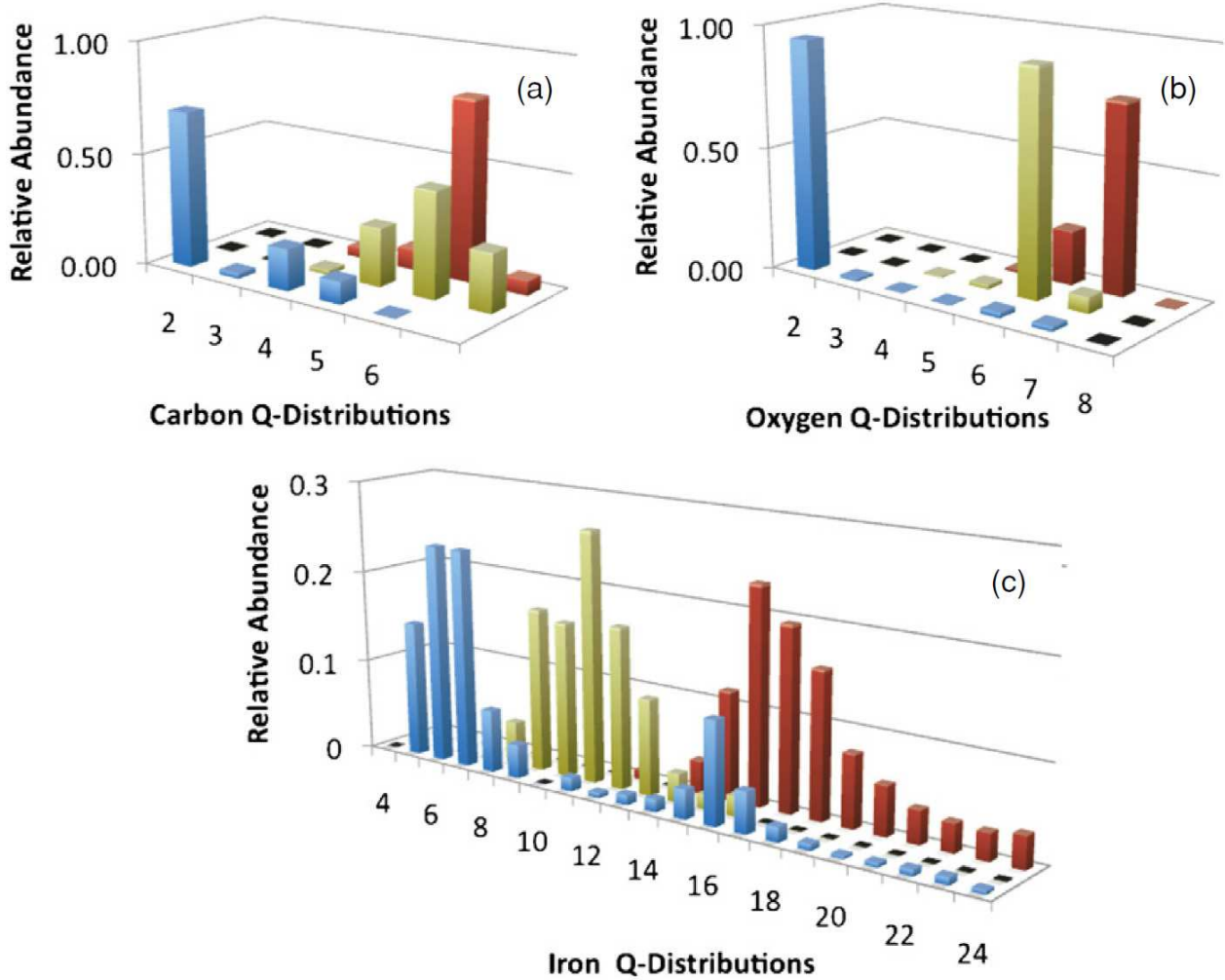}
\caption{Charge state distributions of C, O, and Fe within a cold ICME component (blue), a hot ICME component (red), and normal solar wind (green) \cite{lepri2010}.}
 \label{Fig5}
\end{figure*}

\subsection{The statistical results of ionic charge states within ICMEs}\label{section1}
In this subsection, we introduce a comprehensive statistical study about the charge states of C, O, and Fe within 215 ICMEs from 1998 to 2011 conducted by Owens \cite{owens2018}. The charge state ratios of ion number densities such as C$^{6+}$/C$^{5+}$ and O$^{7+}$/O$^{6+}$ are often used to differentiate the coronal source regions of solar wind \cite{zhaoliang2009,lepri2013,zhaoliang2014,zhaoliang2017apj}. Different with the charge state ratios of C and O, the $<$Q$_{Fe}$$>$ is regarded as a sensitive tracer of electron temperatures and can measure the plasma evolutionary properties in the high corona \cite{lepri2001,lepri2004,lepri2013}.

Owens \cite{owens2018} first divides ICMEs into two groups, i.e., MCs (97 samples) and non-cloud ICMEs (118 samples). Except the average value of charge states within each ICME, he also considers the ICME time profiles of charge states as displayed in the top panels of Figure 6, where the t$_{LE}$ and t$_{TE}$ denote the times when the ICME ejecta leading and trailing edges are detected by ACE, respectively. The duration of each ICME is normalised to one day, and one day before t$_{LE}$ and after t$_{TE}$ is also shown. For fast ICMEs with interplanetary shocks, the t$_{LE}$-1 to t$_{LE}$ interval included the ICME sheath region. The red lines and the pink shaded regions in the top panels show the median and 1-sigma range for MCs, and the blue lines and the blue shaded regions present the same for non-cloud ICMEs.

Figures 6(a) and (b) show that the C$^{6+}$/C$^{5+}$ has little overall enhancement within both MCs and non-cloud ICMEs, while the O$^{7+}$/O$^{6+}$ ratio increases obviously within ICMEs, especially within MCs. Besides, the carbon charge-state profiles show an enhancement near the trailing edge of MCs, while the O$^{7+}$/O$^{6+}$ ratio exhibits a weak decline through MCs. As mentioned, the freeze-in altitudes of C and O charge states are similar in the corona, so they are expected to possess a similar behavior, instead of different ones. This difference might be correlated with the so-called anatomy of depleted ICMEs \cite{kocher2017}, and the similar phenomenon also exists in the background solar wind \cite{zhaoliang2017}, which challenge our understanding of the freeze-in process. The $<$Q$_{Fe}$$>$ also increases obviously within ICMEs as shown in Figure 6(c), consistent with previous studies \cite{lepri2001,lepri2004,song2016}. Figures 6(b) and (c) show that the enhancement in O$^{7+}$/O$^{6+}$ ratio for MCs relative to non-cloud ICMEs is larger than the equivalent for $<$Q$_{Fe}$$>$. This indicates that the preferential heating of MC plasma mainly occurs in the low corona \cite{owens2018}.

\begin{figure*}[t]
\centering
\includegraphics[width=0.95\textwidth]{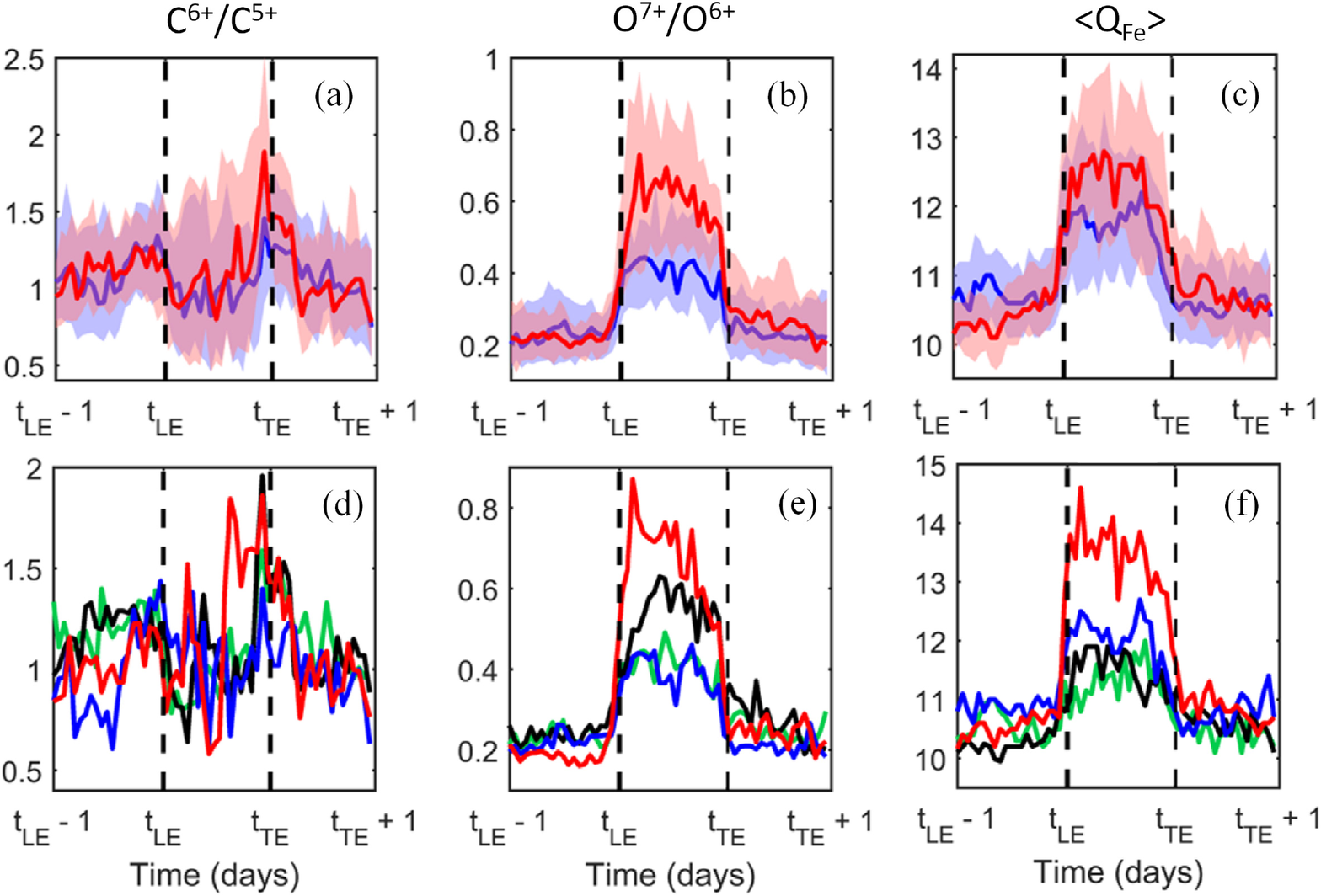}
\caption{A charge state analysis of 215 ICMEs observed by ACE from 1998 to 2011. The ICME ejecta leading and trailing edges are depicted with vertical dashed lines. The ICME duration has been normalised to 1 day. Top panels: red and blue lines present the median for MCs (97 events) and non-cloud ICMEs (119 events), respectively. Shaded regions indicate 1-sigma range. Bottom panels: red and black lines present the median for fast (45 events) and slow (52 events) MCs, respectively. Blue and green lines, for fast (54 events) and slow (64 events) non-cloud ICMEs \cite{owens2018}.}
\label{Fig6}
\end{figure*}

Owens \cite{owens2018} further stratifies the samples by the average speed of ICMEs with a threshold being 450 km s$^{-1}$, and plots the similar time profiles for fast MCs (red, 45 samples), slow MCs (black, 52 samples), fast non-cloud ICMEs (blue, 54 samples) and slow non-cloud ICMEs (green, 64 samples) in the bottom panels of Figure 6, where the uncertainty bands are omitted for clarity. Figure 6(d) shows that the C$^{6+}$/C$^{5+}$ ratio increases obviously within fast MCs as compared to the other groups. Fast MCs also exhibit obviously elevated O and Fe charges states compared to the other groups as shown in Figures 6(e) and (f), which is consistent with other statistics. For example, Lynch et al. \cite{lynch2003} conducted a survey including 56 MCs detected by the ACE close to the maximum period of solar cycle 23. Their fast events exhibit a small increased average profile of O$^{7+}$/O$^{6+}$, and a much stronger profile of Fe$^{16+}$/Fe$_{total}$ compared to the slow cases. Huang et al. \cite{huangjin2020} also reported that fast MCs have enhanced mean charge states of Fe, O, Si, and Mg as compared to the slow MCs. All of the statistical studies give the consistent result. Another feature is that the O$^{7+}$/O$^{6+}$ ratio within slow MCs is obviously higher than the fast non-cloud ICMEs, however, the $<$Q$_{Fe}$$>$ within slow MCs is lower than the fast non-cloud ICMEs.

The above results imply that the freeze-in processes within CMEs are very complex, and the interpretation of charge states is not a straightforward task \cite{boe2018}. For instance, Feng et al. \cite{fengxuedong2018} found that the cold materials within ICMEs can show lower average charge states simultaneously for C, O, Mg, Si, and Fe ions compared to those in the preceding solar wind, and also can contain low charge states of only C with the other ions showing higher average charges compared to those in the preceding solar wind. We suggest to analyze the formation and eruption process of MFRs through the Fe charge states, which are usually enhanced by continual heating through magnetic reconnection in an extended coronal height, different with the C and O charge states that are mainly influenced by heating in the low corona \cite{gruesbeck2011,lepri2012,song2016,owens2018}.

\subsection{The $<$Q$_{Fe}$$>$ distributions within MCs}\label{section1}
Iron is an astrophysically abundant heavy element. Researchers have made important progress on the knowledge of ionization and recombination rates of iron, and its ionic fractions at ionization equilibrium can be calculated and compared with the observational data \cite{arnaud1992}. Lepri et al. \cite{lepri2001} showed that at least 92\% of enhanced ratios of Fe$^{16+}$/Fe$_{total}$ persisting for no less than 20 hours relate to ICMEs near 1 AU. While the observations demonstrated that 63\% of ICMEs possess high Fe charge states. This implies that the Fe ions with high charge states are an excellent sufficient signature for identifying ICMEs but not a necessary condition. The near 100\% correlation of high $<$Q$_{Fe}$$>$ with ICMEs makes them to be a reliable identifier of ICMEs \cite{lepri2004,jianlan2018}. In addition, O$^{7+}$/O$^{6+}$ $>$ 0.7 also acts as an ICME identifier \cite{reinard2001}.

To study the MFR formation and eruption process by means of Fe charge states, Song et al. \cite{song2016} conducted a statistical study on the $<$Q$_{Fe}$$>$ distributions within 96 MCs for solar cycle 23. Following Lepri et al. \cite{lepri2004}, the $<$Q$_{Fe}$$>$ beyond/below 12+ is taken as high/ordinary charge state. According to the criterion, 48 MCs contain the high charge state, and the other 48 cases do not, which is in accordance with the previous ICME percentage related with the high Fe charge states \cite{lepri2001,lepri2004}. The histograms in Figure 7 show the yearly total MC numbers, and the red/blue portions correspond to the events with/without high $<$Q$_{Fe}$$>$. The black line presents the yearly sunspot numbers. Figure 7 exhibits that the MCs with high $<$Q$_{Fe}$$>$ can appear in the rising (1998-1999), maximum (2000), and declining (2001-2006) phases of solar cycle 23, while no high $<$Q$_{Fe}$$>$ MCs are detected during this solar minimum (2007-2009).

\begin{figure}[H]
\centering
\includegraphics[width=0.48\textwidth]{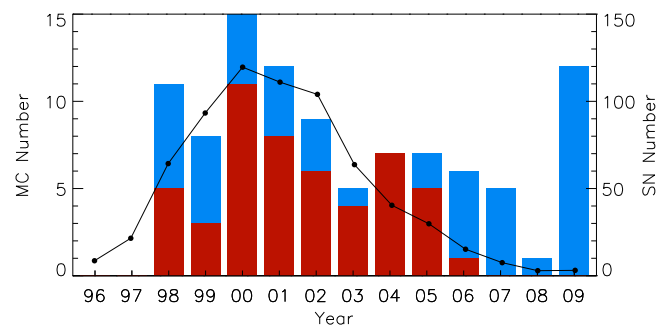}
\caption{Yearly MC numbers in solar cycle 23. The red/blue portion of each bar corresponds to the MCs with/without the high $<$Q$_{Fe}$$>$. The black line presents the yearly sunspot numbers \cite{song2016}.}
 \label{Fig7}
\end{figure}

After inspecting the $<$Q$_{Fe}$$>$ time profiles of 92 MCs, Song et al. \cite{song2016} found that the events can be divided into four groups according to their different features as shown in Figure 8, where the charge-state-distribution map of Fe (top part) and the profile of $<$Q$_{Fe}$$>$ (bottom part) are presented in each panel. In Group A (11 events), the $<$Q$_{Fe}$$>$ presents a bimodal profile with both peaks beyond 12+. Group B (4 events) shows a unimodal profile of $<$Q$_{Fe}$$>$ with peak over 12+. In Group C (29 events) and Group D (48 events), the $<$Q$_{Fe}$$>$ remains beyond and below 12+ along the ACE's trajectory throughout the MC, respectively.

\begin{figure*}[t]
\centering
\includegraphics[width=0.95\textwidth]{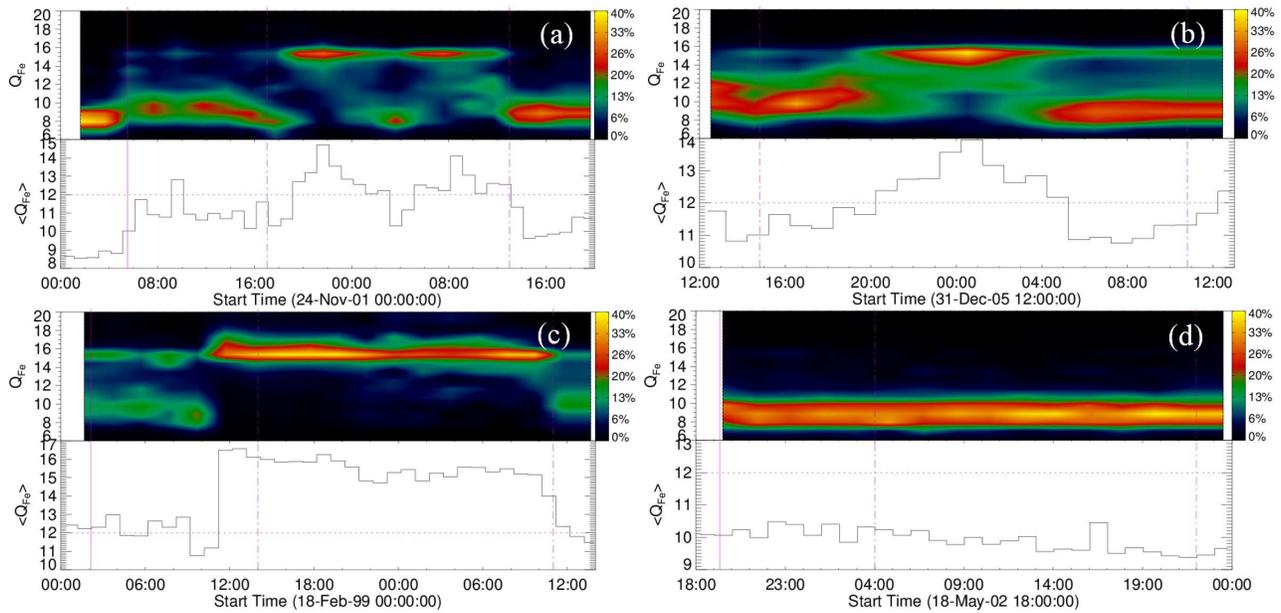}
\caption{The map of Fe charge-state-distribution and the profile of $<$Q$_{Fe}$$>$ (top and bottom in each panel) for each group. (a) bimodal distribution with both peaks beyond 12+. (b) unimodal distribution with the peak beyond 12+. (c) and (d) the $<$Q$_{Fe}$$>$ maintains beyond and below 12+ along the whole trajectory within the MCs, respectively. The vertical solid line and the two vertical dotted-dashed lines denote positions of the ICME shock and two boundaries of the MC, respectively. The horizontal dotted line in each panel depicts the $<$Q$_{Fe}$$>$ level of 12+} \cite{song2016}.
 \label{Fig8}
\end{figure*}

To infer the CME information based on the above in situ observations, one should know how the elevated Fe charge states are produced and made into CMEs. To produce a $<$Q$_{Fe}$$>$ beyond 12+, the plasma should contain Fe ions with a charge range, including both higher and lower than 12+. There will exist many Fe ions with charges states up to and above 16+, which requires that the surrounding plasma must be at temperatures around 5 MK or higher \cite{arnaud1992,lepri2004}. The plasma temperature in flare region can be in excess of 10 MK, and the charge states within ICMEs exhibit a moderate correlation with the flare magnitude, which indicates that the enhanced charge states within ICMEs are generated by the flare-related heating in the corona \cite{reinard2005}. Lepri et al. suggested if magnetic connectivity exists between the CME and the flare region, the plasma with elevated Fe charge states can enter into the CME along the field lines, while no such connectivity will result in no high $<$Q$_{Fe}$$>$ existing within the CME \cite{lepri2004}. Song et al. proposed an alternative scenario \cite{song2016} base on the CME model \cite{forbes1996,lin2000}, in which the magnetic reconnections occurred along current sheets beneath CMEs are able to create high temperatures \cite{forbes1996,lin2000,ciaravella2013}, and the high/ordinary charge state Fe ions are generated in the current sheets with high/ordinary temperature \cite{ciaravella2013}. As the reconnection continues toward the higher altitude, the MFR could be added with more layers that consist of the reconnected field lines. Then the plasma in the current sheet with/without high Fe charge states could be made into the MFR structure along with the field lines, giving rise to the morphology resembling the onion-layer \cite{song2016}.

Song et al. \cite{song2016} described their scenario with schematics as shown in Figure 9 based on the above CME model. The left panels exhibit the situation with a pre-existing MFR as depicted with a yellow circle, in which the blue represents the $<$Q$_{Fe}$$>$ is below 12+. The purple circle denotes the MC boundary. If the temperature of the reconnection region along the current sheet is high, the $<$Q$_{Fe}$$>$ of newly formed MFR portion will be beyond 12+ (indicted with red) as shown in Figure 9(a1). As the ionic charge states freeze-in before the ions leave the corona \cite{rakowski2007,ko2010}, the MC will contain an ordinary/high ionized center/shell. Therefore, a bimodal profile like Figure 8(a) will be observed when the spacecraft passes through one MC along the upper green arrow. On the contrary, when the spacecraft crosses the shell portion along the lower green arrow, a profile like Figure 8(c) would be detected. Figure 9(a2) describes the situation that the temperature of the reconnection region is high at first, while decreases to ordinary subsequently during the MFR growth. Then the MC would possess a high-ionization-state inner shell and an ordinary-ionization-state outer shell. Thus the observed $<$Q$_{Fe}$$>$ profiles can correspond to Figures 8(a) or (b), depending on the trajectory. Figure 9(a3) represents the situation that the current sheet temperature is ordinary (not high enough to generate significant high charge states) during the whole MFR growth. Thus no high charge stated Fe are generated to fill in the MFR, which always results in the profile as Figure 8(d).

The right panels show the situation when the MFR is formed completely during the CME eruption, thus only a purple circle is presented to depict the MC boundary. Correspondingly, Figures 9(b1), (b2), and (b3) represent situations in which plasma temperature within the current sheet is high, first high then ordinary, and ordinary during the MFR growth, respectively. It is not difficult to understand that three $<$Q$_{Fe}$$>$ profiles can be observed, except the bimodal shape in Figure 8(a). Therefore, their analysis proposes that the MFR has existed before the CME eruption when the in situ measurements of the corresponding ICME exhibit the bimodal $<$Q$_{Fe}$$>$ profile. In one word, the observed $<$Q$_{Fe}$$>$ profiles depend on the physical properties of the MFR and the magnetic reconnection in the solar atmosphere, as well as the spacecraft trajectory throughout the MC.

\begin{figure*}[t]
\centering
\includegraphics[width=0.9\textwidth]{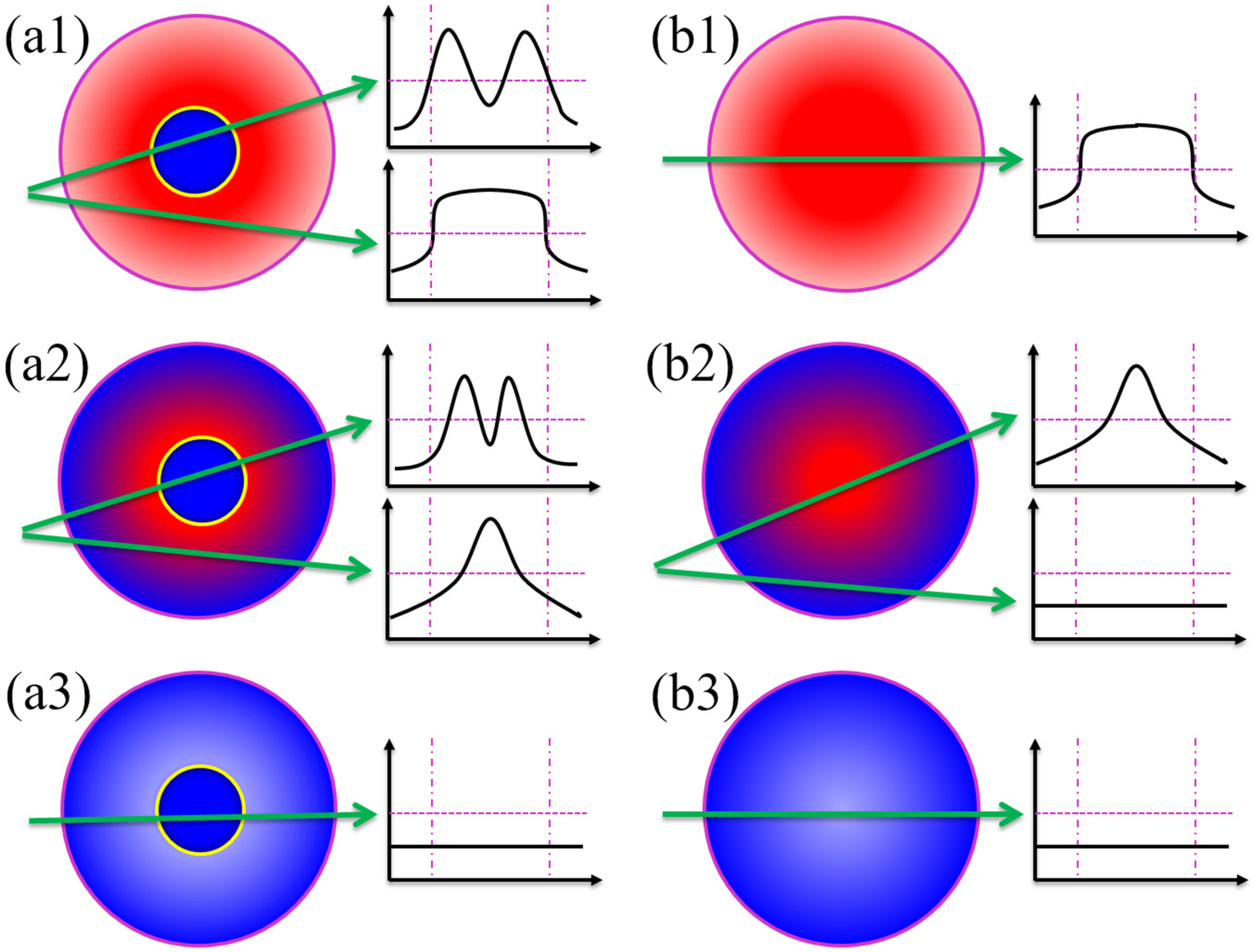}
\caption{Schematic drawings for explaining the $<$Q$_{Fe}$$>$ profiles measured within MCs. Red/blue represents the $<$Q$_{Fe}$$>$ beyond/below 12+. The two purple vertical dotted-dashed lines demarcate the MC boundaries, and the horizontal purple dotted lines mark the level of $<$Q$_{Fe}$$>$ being 12+. The arrows in each panel depict the spacecraft trajectory throughout the MC \cite{song2016}.}
 \label{Fig9}
\end{figure*}

In order to validate the above scenario, Song et al. \cite{song2016} searched the source regions for the 11 MCs in Group A and found the sources for 10 MCs. Using H$\alpha$ images of Big Bear Solar Observatory and EUV images of SOHO, they reported that 7 events are correlated with filaments, which implies a pre-existing MFR \cite{rust1994} and supports their scenario. As no high temperature EUV and soft X-ray data for the rest 3 events, they can not deny or confirm the existence of the other proxies of MFR before the eruptions, such as hot channels \cite{zhangjie2012nc} or sigmoidal structures \cite{titov1999}, which can also lead to the bimodal profile \cite{song2015evidence}. Therefore, this scenario receives preliminary support from observations and has been adopted to investigate MCs \cite{wangwensi2017,huangjin2020}. More information about the formation and eruption process of MFRs can be acquired if the trajectory of the spacecraft throughout the MC could be known.

Except the large-scale MCs, there exist a large quantity of small-scale flux ropes in the interplanetary space \cite{huqiang2018,fenghengqiang2020}. The above scenario has also been applied to investigate the source regions and formation mechanisms of small flux ropes. Based on the $<$Q$_{Fe}$$>$ distribution profiles of small flux ropes, Huang et al. suggested that the small flux ropes originating in the solar corona are formed during eruptions, while those originating in the interplanetary space are correlated with the magnetic reconnection \cite{huangjia2018}. Besides, the elemental abundances and charge states of heavy ions are also used to analyze the filaments within non-cloud ICMEs. The statistical study demonstrated that both enhanced helium abundance and low-ionic-charge-state carbon can be identified in about one third of over 100 non-cloud ICMEs \cite{fengxuedong2018,lidongni2020}.

\section{Summary and discussion}\label{section1}
The velocity, density, temperature, and magnetic field of ICMEs change greatly during their transit to the Earth. However, their composition, including the elemental abundances and charge states of heavy ions, are not obviously altered during the interplanetary propagation. The abundances and charge states act as dyes in ICMEs because they carry information about the plasma origin as well as the formation and eruption process of MFRs in the solar atmosphere due to the FIP effect and freeze-in process. In this review, we mainly introduced the statistical results on the composition of ICMEs, as the ICME composition can vary significantly case by case.

The statistics on the abundances of Fe, Mg, Si, S, C, N, Ne, and He relative to O and H shows that the abundances of ICMEs possess a systematic increase for low-FIP elements (Fe, Mg, and Si) compared to both slow and fast solar wind, which means ICME plasmas possess a larger FIP effect than the solar wind \cite{zurbuchen2016}. In the same time, ICMEs, especially the high $<$Q$_{Fe}$$>$ ICMEs, also present a significant increase of Ne/O component compared to the solar wind. The statistics on C, O, and Fe charge states within MCs and non-cloud ICMEs shows that no obvious difference between them for C charge states. However, the O and Fe charges are obviously elevated within MCs compared to non-cloud ICMEs. Further dividing ICMEs by speed shows that the elevated charge states are primarily limited to fast MCs, and slow MCs are close to non-cloud ICMEs \cite{owens2018}. For more implications of the composition statistics, please consult the references \cite{zurbuchen2016,owens2018}.

The composition analysis provides us a powerful tool to investigate some important issues related to CMEs. We demonstrated what we can learn through the abundances and charge states of heavy ions in ICMEs. For example, two types of models are suggested to answer the origin of filament plasma. One claims that the plasma is brought into the corona from chromosphere or photosphere through an injection/evaporation process or a siphon effect. The other one proposes that the coronal plasma condenses to form the filament plasma due to the thermal instability. The in situ elemental analysis on one erupted filament shows its abundances such as Fe/O, Mg/O, and Si/O are close to their corresponding photospheric values, which does not support that the filament plasma originates from the corona \cite{song2017origin}.

For the ionic charge states, Song et al. checked the $<$Q$_{Fe}$$>$ distribution profiles within 92 MCs one by one and divided them into 4 groups with regular profiles \cite{song2016}. In groups A and B, the $<$Q$_{Fe}$$>$ shows the bimodal and unimodal distributions, respectively, with peaks beyond 12+. In groups C and D, the $<$Q$_{Fe}$$>$ remains beyond and below 12+ during ACE's passage through the MC, respectively. According to their scenario \cite{song2016}, the detailed $<$Q$_{Fe}$$>$ profile depends on the trajectory of spacecraft crossing MCs, as well as the physical properties of the MFR and the current sheet near the Sun. The analysis suggests that the bimodal distribution indicates a pre-existing MFR, which has received preliminary support from observations \cite{song2016}.

Several important issues as discussed below should be addressed further for better utilizing the ICME composition to investigate CMEs. To infer the information about the MFR formation and eruption through charge states, it is crucial to model the ionic composition in reconnection current sheets. Some studies have been conducted to theoretically investigate the ionic charge states of CMEs in the low corona \cite{gruesbeck2011,lynch2011,shenchengcai2013}. For instance, Shen et al. \cite{shenchengcai2013} used a simulation of one large-scale current sheet beneath the CME reported by Reeves et al \cite{reeves2010}. The simulation of Shen et al. included the ohmic and coronal heating, the thermal conduction, as well as the radiative cooling in their energy equation. With the simulation results, they performed the time-dependent ionization calculations of the flow in a current sheet and constructed 2D distributions of the ionic charge states for multiple chemical elements, which demonstrated that significant variability exists in their results when the current sheet was modeled using different models, e.g., the equilibrium ionization model versus a non-equilibrium ionization model. The detailed freeze-in process within the current sheet and MFR might be more complex as compared to the solar wind, thus more efforts are needed to improve our ability to model the ionic charge states within CMEs.

Based on Figure 9, further information about CMEs can be inferred if we know the spacecraft passage through MCs. Several models have been developed, which can reconstruct (e.g., the Grad-Shafranov reconstruction model \cite{huqiang2002,huqiang2017}) or fit (e.g., force-free or non-force-free MFR model \cite{lepping1990,wangyuming2015}) the MCs and help to reveal where the spacecraft passes through MCs. If one ICME can be detected simultaneously by several spacecraft with appropriate separations \cite{liuying2008}, it will be helpful to judge the trajectory. However, usually only one spacecraft passes through one MC, thus it is still difficult to conclude whether the spacecraft is close to the MC center or not as the average diameter of MCs near 1 AU is $\sim$0.20 AU \cite{kumar1996,lepping2015b,liuying2005}. Therefore, more efforts are necessary.

Another unavoidable question is the erosion, which means that the outer structure of MCs might be eroded \cite{ruffenach2012,huangjia2018} during the interplanetary propagation. For a comprehensive review on the ICME propagation, please refer to \cite{manchester2017}. Obviously, it also influences the CME study through charge state distributions within MCs, and the direct solution for this question is to conduct in situ measurements of CMEs close to the Sun. This will provide more valuable results as compared to 1 AU. The Parker Solar Probe \cite{fox2016} launched in 2018 August can swoop to within 9 solar radii of the Sun's surface, while its payloads do not provide composition information of heavy ions. The Solar Orbiter \cite{muller2013} launched in 2020 February will approach the Sun to as close as 0.28 AU, which can provide measurements of solar wind composition for heavy elements and greatly facilitate studies of solar wind and solar eruption. Except the Advanced Space-based Solar Observatory that has been scheduled for launch in 2021 or 2022 \cite{ganweiqun2019b}, Chinese solar physicists are proposing several space missions \cite{ganweiqun2019a} to understand the Sun further \cite{xiongming2016,lavraud2016,lin2019,wangyuming2020}. One of them plans to launch a spacecraft that might approach the Sun to as close as $\sim$5 solar radii eventually \cite{lin2019}, greatly reducing the influence of erosion effect when analyzing the ICME composition.


\Acknowledgements{We thank Prof. TIAN Hui for the invitation to write the review paper. We are grateful to the two anonymous referees for their comments and suggestions that helped to improve the original manuscript. SONG Hongqiang thanks Dr. ZHAO Liang for her helpful suggestions. This work is supported by the Shandong Provincial Natural Science Foundation (JQ201710), the NSFC grants U1731102, U1731101, and 11790303 (11790300), as well as the CAS grants XDA-17040507.}









\end{multicols}

\end{document}